\documentclass[10pt]{iopart}

\usepackage{graphicx}
\usepackage{subfigure}
\usepackage{verbatim}
\usepackage{dcolumn}% Align table columns on decimal point
\usepackage{bm}% bold math
\usepackage{color}
\usepackage[toc]{appendix}
\usepackage{stmaryrd}
\usepackage{amsthm}
\usepackage{hhline}
\usepackage{amssymb}
\usepackage{iopams}
\usepackage{cite}
\usepackage{tikz}
\usetikzlibrary{arrows,shapes,calc,matrix}

\expandafter\let\csname equation*\endcsname\relax
\expandafter\let\csname endequation*\endcsname\relax
\usepackage{amsmath}
\usepackage{xstring}
\usepackage{todonotes}

\makeatletter
\newcommand\footnoteref[1]{\protected@xdef\@thefnmark{\ref{#1}}\@footnotemark}
\makeatother

\newcommand\invisiblesection[1]{%
  \refstepcounter{section}%
  \addcontentsline{toc}{section}{\protect\numberline{\thesection}#1}%
  \sectionmark{#1}}
\newcommand{\bra}[1]{\left\langle #1\right|}
\newcommand{\ket}[1]{\left|#1\right\rangle}
\newcommand{\braket}[2]{\left\langle #1|#2\right\rangle}

\newcommand{\pd}{\partial}

\newcommand{\ex}[1]{\exp{\left(#1\right)}}

\newcommand{\bla}{bla\\bla\\bla\\bla\\bla}
\usepackage{fmtcount}

\newcommand{\mrm}[1]{\mathrm{#1}}

\renewcommand{\appendix}{
}

\newcommand{\draftmode}{1}    %to control draft colors below
\newcommand{\notetoself}[1]{\ifnum \draftmode=1 {\color[rgb]{0,0,0.8} [#1]} \fi}  %notes to self in blue when \draftmode==1.  invisible otherwise
\newcommand{\cuttext}[1]{\ifnum \draftmode=1 {\color[rgb]{0,0.5,0} [#1]} \fi}  %cut out text in green when \draftmode==1.  invisible otherwise
\newcommand{\warntext}[1]{\ifnum \draftmode=1 {\color[rgb]{0.9,0.6,0} #1} \else {#1} \color{black} \fi}
\newcommand{\aref}[1]{{Appendix~\hyperref[#1]{A}}}
\newcommand{\bref}[1]{{Appendix~\hyperref[#1]{B}}}
\usepackage[colorlinks=true,citecolor=blue,linkcolor=blue,urlcolor=blue]{hyperref}
\usepackage{doi}
\usepackage{cleveref}
\begin{document}

%\title{Energy-Efficient Quantum Gates}
\title{From quantum speed limits to energy-efficient quantum gates}

\author{Maxwell Aifer\textsuperscript{1,3}  and Sebastian Deffner\textsuperscript{1,2}}
\address{$^1$Department of Physics, University of Maryland, Baltimore County, Baltimore, MD 21250, USA}
\address{$^2$Instituto de F\'{i}sica `Gleb Wataghin', Universidade Estadual de Campinas, 13083-859, Campinas, S\~{a}o Paulo, Brazil}
\address{$^3$Author to whom any correspondence should be addressed.}
\ead{maifer1@umbc.edu}

\begin{abstract}
While recent breakthroughs in quantum computing promise the nascence of the quantum information age, quantum states remain delicate to control. Moreover, the required energy budget for large scale quantum applications has only sparely been considered.  Addressing either of these issues necessitates a careful study of the most energetically efficient implementation of elementary quantum operations. In the present analysis, we show that this optimal control problem can be solved within the powerful framework of quantum speed limits. To this end, we derive state-independent lower bounds on the energetic cost, from which we find the universally optimal implementation of unitary quantum gates, for both single and $N$-qubit operations.
\end{abstract}

\section{Introduction}

The promise of exploiting quantum features in already existing or near-term technologies \cite{Sanders2017,Preskill2018quantum,Otgonbaatar2021IEEE,domino2021trains} for useful applications has led to increasing funding of quantum initiatives and agendas \cite{Raymer2019QST,Riedel2019QST,Yamamoto2019QST,Sussman2019QST,Roberson2019QST},  with global investment in quantum computing research expected to grow by over 10\% annually through 2027 \cite{quantum_spending}. Remarkably, during the same period, global spending on quantum computing products is predicted to increase by over 50\% annually \cite{quantum_spending}. In anticipation of this growth, it seems important to clearly define the ``quantum promise'', both with regards to its impact on society \cite{Aiello2021QST,Roberson2021QST} and with regards to its physical and technological ramifications. 

Mathematically motivated progress towards an answer to this question has been made in the field of computational complexity, where problems are classified by the amount of resources necessary to solve them on classical and quantum computers. It has been shown, for instance, that an integer can be factored using a quantum computer in an amount of time which is upper bounded by a polynomial function of the number of digits the integer has \cite{shor_polynomial-time_1999}.  To the very best of our knowledge,  to date no algorithm has been published which runs on a classical computer and solves the same problem in polynomial time. Another example of a theoretical ``quantum speedup'' is a quantum algorithm for training perceptrons, a fundamental task in machine learning, which requires an amount of time scaling with the square root of the number of data points \cite{wiebe_quantum_2016}. The best known classical algorithm requires an amount of time which is linear in the number of data points \cite{wiebe_quantum_2016}.

While the asymptotic scaling of resource usage with problem size on quantum computers is increasingly well understood \cite{auffeves2021}, there has been less emphasis placed on calculating the coefficients which determine this dependence exactly; that is, it may be known that an algorithm can be executed in an amount of time which is proportional to a particular function of the input size, but the constant of proportionality is seldom known precisely. This is partly because the coefficients differ from one machine to another, and are therefore not fundamentally properties of the algorithm \cite{arora_computational_2009}. An additional wrinkle is that the execution time is often estimated by reference to the ``gate complexity'' \cite{Koike2010PRA}, which is the number of gates needed to implement an algorithm in the quantum circuit model. Depending on which set of gates one takes to be the smallest building blocks of the algorithm (called ``elementary operations''), the coefficients may vary.

 The importance of these coefficients is underscored by current efforts to create exascale classical computers around the world \cite{moss_race_2019}. These systems will make new computations feasible, not by decreasing the number of operations necessary to complete the task, but by dramatically reducing the time cost per operation. It is also informative to compare the current monetary cost of elementary operations on classical and quantum computers. Recent benchmarking of the Amazon Web Services (AWS) computing platform shows that the platform charges about $4 \times 10^{-13}$ cents per classical floating point operation, whereas a single quantum circuit evaluation costs $1$ cent on an AWS-owned Rigetti machine \cite{akan_performance_2010,noauthor_amazon_2022}. Given such a pricing scheme, while it is certainly true that there must be some problem size above which an $O(\sqrt{N})$ quantum algorithm will be more cost-effective than an $O(N)$ classical algorithm, it is far from obvious that this problem size is exceeded in practical use-cases.

It is therefore desirable to find a measure of the cost of running a quantum algorithm which can be precisely quantified, independent of the specific hardware used. As stated earlier, the execution time of an algorithm is hardware-dependent, but this is not to say that the  execution time can be decreased at will without incurring some other cost \cite{auffeves2021}. In particular,  quantifying the amount of energy required to operate quantum technologies is crucially important \cite{auffeves2021}.  It is interesting to note that quantifying energy usage for computing is an inherently important area of research. The issue of energy-efficiency in computing has recently received increased attention due to the high stakes involved for Earth's climate. For instance, regulators have announced that goals set in the Paris Climate Agreement are not likely to be reached unless bitcoin mining is banned across all of Europe, due to the large energy consumption of mining algorithms \cite{bateman_europe_2021}.  Another concerning revelation is that ``training a single AI model can emit as much carbon as five cars in their lifetimes"\cite{hao_training_2019}.  

Thus,  as a first and foundational step towards addressing this important problem,  we pose the question: ``What is the minimum amount of energy needed to implement a quantum gate in a set amount of time?''  We will answer this question for general quantum gates of arbitrary dimension, and will provide an explicit expression for a protocol which attains this minimum energetic cost in the form of a Hamiltonian $H$. We find that in general the minimum energetic cost is nonzero, and depends on the particular gate being implemented. Moreover, this lower bound cannot be violated by any machine.  Since the elementary operations in the circuit model are individual gates, our findings can be used to efficiently implement a quantum algorithm consisting of many gates. Our analysis is rooted in an active field of modern research that is focused on so-called quantum speed limits (QSLs) \cite{Pfeifer1995,Frey2016,deffner_quantum_2017,campaioli_tight_2019}.

These QSLs are a collection of theorems which originate in mathematically rigorous treatments of Heisenberg's uncertainty relation for energy and time \cite{Heisenberg1927}.  While originally only formulated for isolated,  undriven dynamics \cite{tamm_uncertainty_1991,margolus_maximum_1998,giovannetti2003,levitin_fundamental_2009}, in recent years QSLs have been generalized to controlled dynamics \cite{Jones2010,Deffner2013JPA} and open quantum systems \cite{taddei2013,deffner2013PRL,DelCampo2013,mirkin2016,funo2019}.  For a comprehensive review we refer to the literature \cite{Pfeifer1995,Frey2016,deffner_quantum_2017}.  In principle,  studying the QSL has given rise to two fundamentally different questions \cite{Deffner2014JPB}, namely either to quantify the minimal time a quantum system needs to evolve between distinct states \cite{caneva2009,Hegerfeldt2013,poggi2013,Hegertfeldt2014,Xi2016,wakamura_general_2020}, or to bound the maximal rate with which quantum states can evolve \cite{Pires2016,Deffner2017NJP,FogartyPRL2020,Poggi2021PRXQ}.  While either version of the QSL can be exploited to assess the time cost and energy cost of a quantum computation \cite{Campbell2017,poggi2019,Deffner2020PRR},  previous formulations are limited by the fact that the QSL  has been considered as an inherent property of the evolving state -- and not of the Hamiltonian generating the dynamics.

Thus,  our analysis starts with a derivation of a new QSL, which is independent of the evolving state, and is fully determined by the eigenvalues of the governing Hamiltonian. This is essential to quantify the cost of a quantum gate,  independent of the input and output states.  To this end, several new quantities need to be defined.  Each quantum gate has an associated ``shortest covering arc length'', which is a nonnegative scalar that measures how difficult it is to implement the gate.  We prove a theorem which states that the minimum energetic cost of implementing a gate is proportional to its shortest covering arc length.  Moreover,  we also derive a lower bound on the closely related ``energy-time phase volume'' of evolution.  As a main result, we formulate a ``recipe'' to design the most energetically efficient implementation of single as well as $N$-qubit gates.

\section{Speed Limits and Energetic Cost}
\label{section-2}

We begin with a mini-review of notions and notations. This will motivate and inform how to quantify the energetic cost of any (unitary) quantum operation. Ideally, we would like a quantum computer to go between distinguishable states as quickly as possible and while using as little energy as possible.  Mathematically, the question may be posed as follows: suppose that at time $t=0$ the system is in a known state $\ket{\psi_0}$. The system is assumed to be isolated from the rest of the universe, and has a constant Hamiltonian $H$, and hence the state of the system evolves according to the Schr\"odinger equation, $i \hbar \pd_t \ket{\psi}=H\ket{\psi}$. Subject to these assumptions, what is the smallest possible time $\tau$ such that $\braket{\psi(0)}{\psi(\tau)}=0$?

The minimal time to reach an orthogonal state is then given by the Mandelstamm-Tamm (MT) inequality \cite{tamm_uncertainty_1991},$ \tau \geq \hbar \pi/2\Delta E$,  where $\Delta E$ is the standard deviation in energy of the initial state. A seemingly independent bound on the orthogonalization time  is given by the Margolus-Levitin (ML) inequality \cite{margolus_maximum_1998},  $\tau \geq \hbar\pi/2\left(\langle E\rangle - E_0\right)$,  where $\langle E \rangle =\bra{\psi} H \ket{\psi}$ and $E_0$ is the ground state energy.  However, it has been noted that the ``unified" bound is tight \cite{levitin_fundamental_2009}, and hence the QSL time for orthogonal states can be written as
\begin{equation}
\label{unified}
\tau \geq  \tau_\mrm{QSL}\equiv \max \left\{ \frac{\hbar\pi/2}{\Delta E} , \frac{\hbar\pi/2}{\langle E \rangle - E_0}\right\}.
\end{equation}

Many, but not all, computations have the property that they leave the computer in a state which can be distinguished from its original state. Two pure states $\ket{\psi_1}$ and $\ket{\psi_2}$ can be perfectly distinguished from each other if and only if they are orthogonal \cite{Wootters1982,nielsen_quantum_2000}.  However,  in the study of QSLs it is often instructive to consider arbitrary angles \cite{Bhattacharyya1983}, and hence less then perfect distinguishability. The angle between two pure states is simply defined as
\begin{equation}
\mathcal{B}(\psi_1,\psi_2)=\arccos{\left|\braket{\psi_1}{\psi_2}\right|}.
\end{equation}
Choosing some angle $\theta$, we then ask what is the smallest $\tau$ for which $\mathcal{B}(\psi_0,\psi(\tau))=\theta$. A tight bound is given by a straightforward generalization of Eq. (\ref{unified}) \cite{deffner_energytime_2013},
\begin{equation}
\label{unified-angle}
\tau \geq \tau_\mrm{QSL}\equiv \max \left\{ \frac{\hbar\theta}{\Delta E} , \frac{\hbar\theta
}{\langle E \rangle - E_0}\right\}.
\end{equation}
However, computing the such defined $\tau_\mrm{QSL}$ requires knowledge of both the initial and final states.

More typically for a computer,  the initial and final states are unknown. In this case it is not possible to evaluate $\Delta E$ and $\langle E \rangle-E_0$. However, these quantities can be upper bounded using the fact that the energy eigenvalues are necessarily bounded for a finite-dimensional system. Consider $N$ energy eigenvalues that are sorted in increasing order ($E_0 \leq E_1 \leq \dots E_{N-1}$), and we assume without loss of generality that $E_0\geq 0$\footnote{If the ground state energy was negative, $E_0<0$, we could always ``shift'' the energy spectrum by including a simple additive term in the Hamiltonian}.  Now observe that since $\langle E \rangle$ is at most $E_{N-1}$, the least upper bound on $\langle E \rangle - E_0$ is $E_{N-1}-E_0$. This leads to the following ``state-independent" ML inequality
\begin{equation}
\label{best-case-ML}
\tau \geq  \frac{\hbar\theta}{E_{N-1}-E_0}.
\end{equation}
 We proceed similarly with the MT result. A theorem known as Popoviciu's inequality \cite{bhatia_better_2000} asserts that $\Delta E \leq (E_{N-1}-E_0)/2$, and so the corresponding state-independent MT inequality is
\begin{equation}
\label{best-case-MT}
\tau \geq \frac{2\hbar\theta}{E_{N-1}-E_0}.
\end{equation}
Equation (\ref{best-case-MT}) is our first main result. As desired, this speed limit can be computed from the Hamiltonian alone, without needing to know the initial state. Whereas Eq. (\ref{unified-angle}) states that a \emph{specific} state cannot travel an angle $\theta$ in time less than $\tau$, Eq. (\ref{best-case-MT}) says that \emph{no} state can travel an angle $\theta$ in time less than $\tau$. Equation (\ref{best-case-ML}) is also a valid speed limit, but since it is less restrictive than Eq. (\ref{best-case-MT}), it is not practically useful. Equation (\ref{best-case-MT}), in turn, is less restrictive than Eq. (\ref{unified-angle}); however, it has the advantage that it can be computed when the system's state is not specified.

\begin{figure}
\label{area-graph}
\centering
\includegraphics[width=.8\textwidth]{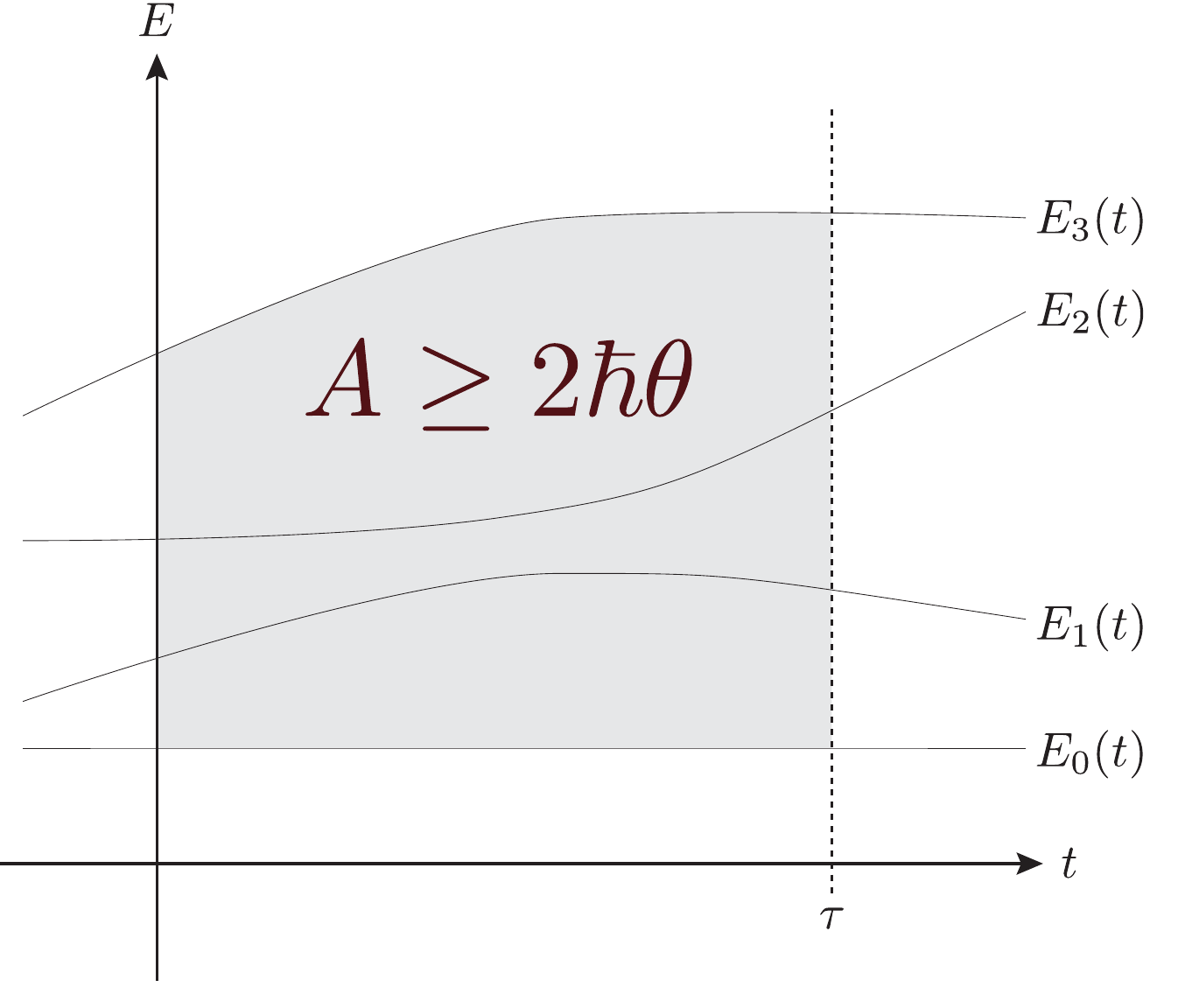}
\caption{\label{fig:area} The area contained in the shaded region is bounded below by the distinguishability between the initial and final states, compare Eq.~\eqref{QSL-area}.}
\end{figure}

A generalization of (\ref{best-case-MT}) is obtained by allowing for a time-dependent Hamiltonian $H(t)$. We assume that for some large integer $M$, if the interval $(0,\tau)$ is subdivided into $M$ segments of equal length, then the Hamiltonian is approximately constant within each segment. Applying Eq. (\ref{best-case-MT}) in each segment gives
\begin{equation}
\frac{\tau}{M} (E_{N-1}(t_j)-E_0(t_j)) \geq 2\hbar\,\delta \theta_j,
\end{equation}
Where $t_j=j \tau/M$ and $\delta \theta_n = \mathcal{B}(\psi(t_n),\psi(t_{n+1}))$
Summing over all segments, we obtain
\begin{equation}
\sum_{j=1}^M \frac{\tau}{M} (E_{N-1}(t_j)-E_0(t_j)) \geq 2\hbar \sum_{j=1}^M \delta \theta_j.
\end{equation}
Let $\theta$ again be the angle between the initial and final states, i.e. $\theta = \mathcal{B}(\psi_0,\psi(\tau))$. It is easy to see that $\mathcal{B}(\cdot,\cdot)$ is a distance function which satisfies the triangle inequality \cite{nielsen_quantum_2000}. Therefore $\theta$ cannot exceed the sum of the angles $\delta \theta_n$. Using this fact and taking the limit of $M \to \infty$ gives
\begin{equation}
\label{QSL-area}
A\equiv\int_0^\tau dt\,( E_{N-1}(t)-E_0(t))\geq 2\hbar\theta.
\end{equation}
Equation (\ref{QSL-area}) states that when the energy eigenvalues are plotted as functions of time, the area contained between the graphs of the largest and smallest eigenvalues must be at least $2\hbar \theta$, see Fig.~\ref{fig:area} for an illustration. It should be noted that in the single qubit case $2\theta$ is the angle between the initial and final states on the Bloch sphere. Due to this graphical interpretation, we refer to the quantity on the left hand side of Eq. (\ref{QSL-area}) as the energy-time phase volume, and denote it by $A$.

Equations (\ref{best-case-MT}) and (\ref{QSL-area}) can be used to derive corresponding inequalities in terms of the Schatten-$p$-norm of the Hamiltonian, 
\begin{equation}
\label{schatten-p}
\|H\|_p = \left( \sum_{n=0}^{N-1}|E_n|^p \right)^{1/p}.
\end{equation}
It is clear from the definition (\ref{schatten-p}) that $\|H\|_p \geq  |E_n|$ for all $n$, which implies that $E_{N-1} - E_0 \leq 2\|H\|_p$ for all $p$. 
\begin{figure}
\centering
\includegraphics[width=.8\textwidth]{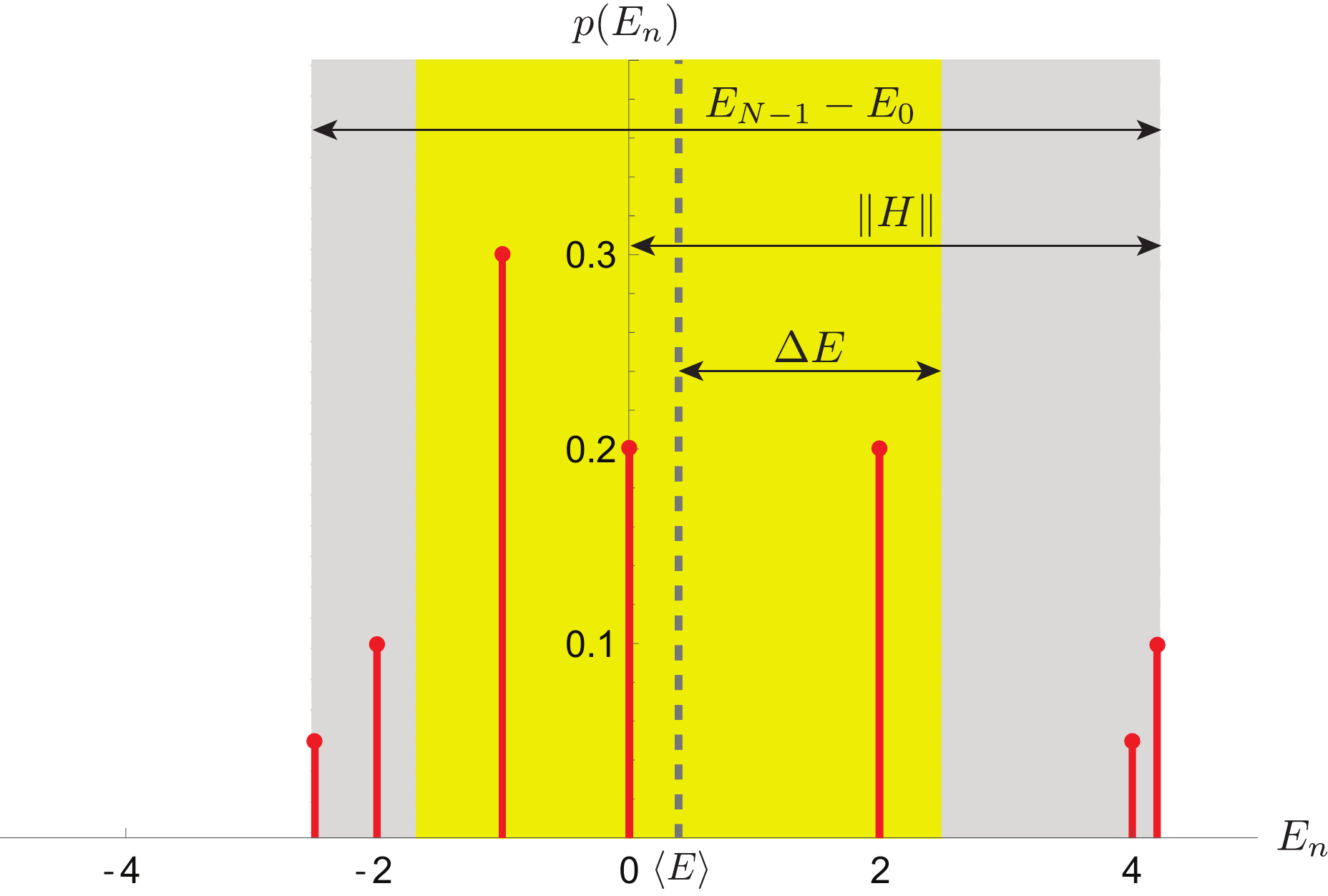}
\caption{Comparison of three different estimates of energy spread for a state with an arbitrarily-chosen energy distribution. In red we depict the energy distribution,  the yellow shaded area corresponds to the standard deviation, and the grey shaded area is the ``maximal'' spread of energy eigenvalues.  Note that the ``maximal spread'' is clearly smaller than twice the operator norm $\|H\|$.}
\label{fig:H-norm-comparison}
\end{figure}
In particular, taking $p$ to infinity gives the operator norm, defined as the maximum of the absolute values of the energy eigenvalues
\begin{equation}
\label{op-norm}
\|H\|=\max \{ |E_0|, |E_1|, \dots, |E_{N-1}|\},
\end{equation}
and for which we still have $E_{N-1} - E_0 \leq 2\|H\|$, see Fig.~\ref{fig:H-norm-comparison} for an illustration. Therefore, from (\ref{best-case-MT}), we immediately obtain the inequality
\begin{equation}
\label{angle-schatten}
\tau \geq \frac{\hbar\theta}{\|H\|_p},
\end{equation}
valid for all $p$, including $p=\infty$. Using Eq. (\ref{angle-schatten}) in the derivation of Eq.~\eqref{QSL-area} we obtain
\begin{equation}
\label{QSL-energetic-cost}
\int_0^\tau  dt\,\|H(t)\|_p\geq \hbar \theta.
\end{equation}

The quantity appearing on the left hand side of Eq. (\ref{QSL-energetic-cost}) generalizes the energetic cost functional first introduced in Ref.~\cite{zheng_cost_2016}, and then used by Abah \etal. \cite{abah_energetic_2019} to assess different methods of shortcuts to adiabaticity \cite{Guery2019RMP}.  It is interesting to note that this quantity does correspond in some cases to the amount of energy required to control a quantum system. For instance, for a qubit in a magnetic field, the energy used to create the field is proportional to this expression.  Therefore, we will henceforth refer to this quantity as simply the ``energetic cost''. Unless otherwise specified,  we set $p=\infty$, which provides a lower bound for all other cases. Interestingly, the same energetic cost appeared in Ref.~\cite{deffner_energetic_2021}, in a quantum analog of Landauer's principle for unitary quantum evolution.

As was shown above, the energetic cost gives a general form of the quantum speed limit. Having dimensions of energy times time, the energetic cost is formally an action \cite{oconnor_action_2021}. When it is minimized using variational methods, the formalism becomes analogous to Lagrangian mechanics, a parallel which is highlighted in Sec.~\ref{section-5}. 

As was explained above, Eq. (\ref{QSL-energetic-cost}) is useful when the initial state of the system is ambiguous, as opposed to Eq. (\ref{unified}), which is useful when the initial state is specified. An important application where the initial state is generally unspecified is the operation of quantum gates. A quantum gate is a unitary operator $U$ acting on the Hilbert space $\mathcal{H}$, which maps each state $\ket{\psi}$ to a state $U \ket{\psi}$. We would like to have a ``gate quantum speed limit'', which describes the energy-time trade-off necessary for operating a quantum gate. However, Eq. (\ref{QSL-energetic-cost}) is still not suitable for this purpose,  since the angle $\theta$ is also undetermined in the case of quantum gates. This is due to the fact that the distinguishability between the initial and final states depends on what the initial state is. In Sec.~\ref{section-3}, we will identify the correct quantity to take the place of $\theta$ for single qubit gates, which will be generalized to gates of arbitrary dimension in Sec.~\ref{section-4}.

 \section{Energy-efficient single qubit gates}
 \label{section-3}
 
Our main objective is now to find the minimum energetic cost of implementing such single qubit gates as well as the corresponding optimal protocol.  To this end, consider a unitary operator $U$ in the form of a $2\times 2$ matrix, and we want to find a Hamiltonian $H(t)$ such that for some time $\tau$, independent of the initial state $\ket{\psi(0)}$, we will have
 \begin{equation}
 \label{gate-condition}
 \ket{\psi(\tau)} = \ex{i\varphi}\,U\ket{\psi(0)},
 \end{equation}
for some undetermined real number $\varphi$.  We allow for an arbitrary overall phase factor $\ex{i\varphi}$, since we consider operations on a single qubit in isolation.  Although $\varphi$ may take any value, its value must be the same regardless of the initial state.  Constructing such Hamiltonians can be done systematically by exploiting techniques from inverse engineering \cite{Santos2017}. For our purposes, we we will call a time-dependent Hamiltonian an \emph{allowed protocol}, if the corresponding dynamics satisfy \eqref{gate-condition}. We are then interested in minimizing the energetic cost functional over the set of all allowed protocols,
\begin{equation}
\label{C-def}
C[H] = \int_0^\tau dt\,\|H(t)\|.
\end{equation}
Any allowed protocol which minimizes $C$ is called an \emph{optimal protocol}. The problem is then to find the minimum cost and at least one optimal protocol, given an arbitrary unitary operator $U$.

\subsection{Recasting the problem in terms of Bloch sphere rotations}

For the single qubit case, the Hamiltonian $H(t)$ takes the form of a $2\times 2$ Hermitian matrix and it can be expanded in the basis which consists of the three Pauli matrices $(\sigma_x, \sigma_y, \sigma_z)$ along with the identity matrix $\mathbb{I}$.  We have
\begin{equation}
\label{H-def}
H(t) = u_0(t) \mathbb{I} + u_1(t) \sigma_x + u_2(t) \sigma_y + u_3(t) \sigma_z\,.
\end{equation}
The instantaneous eigenvalues of this Hamiltonian are then
\begin{equation}
E_{\pm} = u_0(t) \pm \sqrt{u_1(t)^2 + u_2(t)^2 + u_3(t)^2}\,.
\end{equation}
This implies that the operator norm of the Hamiltonian simply reads
 \begin{equation}
 \label{op-norm-from-u}
     \|H(t)\| = |u_0(t)| + \sqrt{u_1(t)^2 + u_2(t)^2 + u_3(t)^2}.
 \end{equation}
It can be shown that varying $u_0(t)$ only affects the overall phase of the state as it evolves (see Appendix \ref{appendix-1}). Therefore for any allowed protocol with nonzero $u_0(t)$, we can obtain a new allowed protocol which has a lower cost by setting $u_0(t)=0$ everywhere. 

Next, we observe that if the wavefunction at time $t$ is $\ket{\psi(t)}$, then a short time later, at time $t+dt$, the wavefunction will be
\begin{equation}
\label{eq:infinitesimal}
 \ket{\psi(t+dt)} =\left(1-\frac{i}{\hbar}\,H(t)dt\right)\ket{\psi(t)} =\left(1-\frac{i}{\hbar}\,\vec{\sigma}\cdot \vec{u}(t)\,dt\right)\ket{\psi(t)} 
 \end{equation}
Equation~\eqref{eq:infinitesimal} is simply an expression of the evolution operator to first order in time. Comparing this with the definition of the infinitesimal rotation $R(\hat{n},d\phi)$ \cite{sakurai_modern_2011}, we immediately recognize
 \begin{equation}
     \ket{\psi(t+dt)} = \hat{R}\left(\hat{u}(t), \frac{2|\vec{u}(t)|}{\hbar}dt\right)\ket{\psi(t)},
 \end{equation}
which means that $\ket{\psi(t+dt)}$ evolves by rotating $\ket{\psi(t)}$ by an infinitesimal angle $2|\vec{u}(t)|/\hbar\,dt$ around the $\hat{u}(t)$-axis.  As always, $\hat{u}(t)\equiv \vec{u}(t)/|\vec{u}(t)|$. It is equivalent to say that the Bloch vector rotates with angular velocity $ \vec{\omega}\equiv 2 \vec{u}/\hbar$. Thus, it follows from Eq. (\ref{op-norm-from-u}) that the operator norm of the Hamiltonian is simply given by $ \|H(t)\| = \hbar|\vec{\omega} (t)|/2$, and we can therefore express the cost functional (\ref{C-def}) as
\begin{equation}
\label{C-from-omega}
C = \frac{\hbar}{2} \int_0^\tau dt \left| \vec{\omega}(t) \right|.
\end{equation}

By Euler's rotation theorem, a sequence of rotations is itself a rotation. Therefore the transformation accomplished by any unitary evolution can be described (up to global phase) by a single rotation of the Bloch sphere about an axis $\hat{n}$ by an angle $\alpha$.  To determine the parameters $\alpha$ and $\hat{n}$ we compare the form of the unitary operator $U$ with the definition of the rotation operator.  Since we are ignoring the overall phase, we may take $U$ to have determinant $1$. In this case $U$ must take the form \cite{tinkham_group_2003}
\begin{equation}
\label{SU2-matrixform}
 U =\begin{pmatrix}
 a & b \\
  -b^*& a^*
 \end{pmatrix},
 \end{equation}
where $|a|^2+|b|^2=1$. A general rotation operator with axis $\hat{n}$ and angle $\alpha$ is given by
 \begin{equation}
 \label{rotation}
R = \begin{pmatrix}
 \cos (\alpha/2)-i n_3 \sin (\alpha/2) & \sin (\alpha/2)(-n_2-i n_1) \\
 \sin (\alpha/2)(n_2-i n_1) & \cos (\alpha/2)+i n_3 \sin (\alpha/2) 
\end{pmatrix}
 \end{equation}
 Identifying the elements of the matrices in Eqs. (\ref{SU2-matrixform}) and (\ref{rotation}), we obtain expressions for the parameters $n_1$, $n_2$, $n_3$, and $\alpha$ in terms of the matrix elements $a$ and $b$. There is still some ambiguity in choosing the parameters $\hat{n}$ and $\alpha$,  since the rotations $R(\hat{n}, \alpha)$ and $R(\hat{n}, \alpha \pm n 2\pi)$ are equivalent for all $n \in \mathbb{Z}$. Similarly, the rotations $R(\hat{n}, \alpha)$ and $R(-\hat{n}, 2\pi - \alpha)$ are equivalent. We eliminate this ambiguity by requiring that $0 \leq \alpha \leq \pi$. Subject to this condition, we have
 \begin{equation}
  \label{alpha-from-U}
 \alpha = 2\arccos(\text{Re}[a])\quad\text{and}\quad \hat{n} = -\frac{1}{\sin(\alpha/2)}
\begin{pmatrix}
\text{Im}[b] \\ \text{Re}[b] \\\text{Im}[a]
\end{pmatrix}.
 \end{equation}
In conclusion, we have translated finding energetically optimal quantum gates to a simple geometric problem on the Bloch sphere.

\subsection{Solution to the single qubit problem}

To find the actual solution of the optimal control problem,  we now need to determine a function $\vec{\omega}(t)$ such that if each point on the Bloch sphere moves with angular velocity $\vec{\omega}$ from time $t=0$ to time $\tau$, then the end result will be a rotation of the Bloch sphere by an angle $\alpha$ about an axis $\hat{n}$. Of all such functions, we must find one which minimizes Eq. (\ref{C-from-omega}), and then compute the minimal value of the cost.

A natural guess, which turns out to be correct, is that the rotation can be implemented with minimal cost by rotating about the axis $\hat{n}$ at a constant rate\footnote{Note that we are searching for the optimal protocol that drives the quantum state along a geodesic of the Riemannian manifold corresponding to the cost functional \eqref{C-def} \cite{Andersson2014JMP,Andersson2014PS,Allan2021quantum,oconnor_action_2021}.}. We will now show that the following protocol minimizes the cost, $\vec{\omega}(t) = \alpha/\tau\, \hat{n}$.  First, it is apparent that this is an allowed protocol since the axis of rotation is $\hat{n}$ and the sphere rotates by an angle $\alpha$ between time $0$ and time $\tau$, so this process results in the correct transformation of each state. By plugging the constant protocol into Eq. (\ref{C-from-omega}) we see that the cost for this protocol simply is $C = \hbar\alpha/2$.

What remains is to show that no other protocol can have a lower cost.  To this end,  consider a vector $\vec{v}_0$ which is perpendicular to $\hat{n}$. Let $\vec{v}(t)$ be the trajectory of the point on the Bloch sphere which starts at $\vec{v}_0$.  After the Bloch sphere has been rotated, the angle between $\vec{v}(\tau)$ and $\vec{v}_0$ will be $\alpha$.  

\begin{figure}
\centering
\includegraphics[width=.8\textwidth]{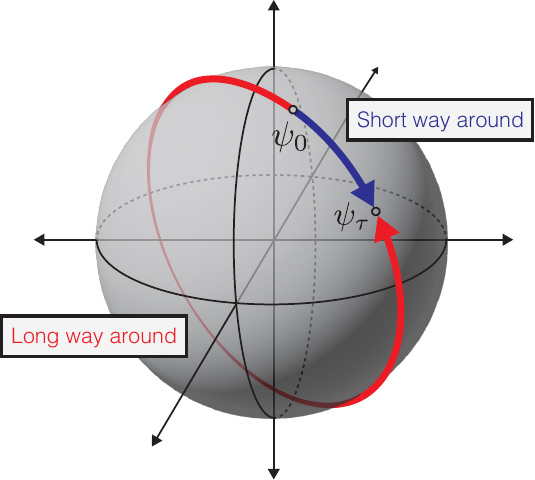}
\caption{Two paths on the Bloch sphere that accomplish the same rotation, but one is longer}
\label{long-vs-short}
\end{figure}

This simple geometric picture then makes the argument almost trivial.  The geodesics on the surface of a sphere are the great circles.   Thus,  the length of the path traveled by the vector during the interval $(0,\tau)$ is at least $\alpha$.  Note that this would not necessarily be true if $\alpha$ were allowed to be greater than $\pi$. The fact that $\alpha$ is required to be less than $\pi$ is equivalent to assuming that the Bloch sphere is rotated ``the short way around'' in order to reach the desired orientation, see Fig. ~\ref{long-vs-short}.

The integral of the speed of the vector $\vec{v}$ on the surface on the Bloch sphere is therefore at least $\alpha$, that is
\begin{equation}
\label{min-speed-int}
\int_0^\tau dt \left|\frac{d\vec{v}}{dt}\right| \geq \alpha.
\end{equation}
Now exploiting the vector identity $\pd_t \vec{v}=\vec{\omega} \times \vec{v}$, we have 
\begin{equation}
\int_0 ^\tau dt\, \left|  \vec{\omega} \times \vec{v}\right| \geq \alpha\,, \quad\text{and thus}\quad \int_0^\tau dt|\vec{\omega}(t)| \geq \alpha
, 
\end{equation}
where we used that $|\vec{\omega}| |\vec{v}| \geq|\vec{\omega} \times \vec{v} |$.  Therefore, we immediately obtain a lower bound on the energetic cost,
\begin{equation}
\label{1-qubit-min-cost}
\int_0^\tau dt\,\|H(t)\| \geq \hbar \alpha/2,
\end{equation}
which is exactly what we set out to prove.

Equation (\ref{1-qubit-min-cost}) can be thought of as a QSL for single qubit gates. This is especially apparent when it is put in a form more similar to the ML and MT inequalities,
\begin{equation}
\label{qsl-2-by-2}
\tau \geq \frac{\hbar \alpha}{2 \langle \|H\| \rangle_t},
\end{equation}
where $\langle \cdot \rangle_t$ denotes time averaging. It is useful to compare this result to other QSL results to better understand how it fits into this body of literature. The MT and ML bounds describe the minimum time necessary to get a system to go between a pair of \emph{known} states. In Sec.~\ref{section-2} we derived a state-independent QSL in Eqs. ~\eqref{best-case-MT}, (\ref{angle-schatten}), and (\ref{QSL-energetic-cost}). These results give the minimum time needed to go between a pair of \emph{unknown} states which are separated by a \emph{known} angle. Finally,  Eq.~\eqref{qsl-2-by-2} is a QSL for arbitrary quantum gates. This bounds gives the minimum time necessary to \emph{implement a quantum gate} on an \emph{unknown} state, such that the system will end up in a state which is separated from the first by an angle which is also unknown.

Comparing Eq. ~\eqref{1-qubit-min-cost} with Eq. ~\eqref{QSL-energetic-cost}, we observe that for a quantum gate the parameter $\alpha/2$ plays  the same role that the angle $\theta$ plays for the transition between two pure states. This can be understood by considering that the angle $\theta$ is equal to half the angle between states on the Bloch sphere. When a sphere is rotated about some axis by an angle $\alpha$, the maximum angle which a point on the sphere goes through is $\alpha$, which implies that
\begin{equation}
\frac{\alpha}{2} = \max_{\psi_0}\left\{ \mathcal{B}(\ket{\psi_0}, U \ket{\psi_0})\right\}.
\end{equation}
Hence, $\alpha$ can be understood as a measure of the difficulty of implementing a given unitary gate,  since it sets the minimally required energetic cost.

\subsection{Pedagogical example: the Hadamard gate}
The Hadamard gate is a single qubit gate, which we write as
\begin{equation}
G= \frac{1}{\sqrt{2}}\begin{pmatrix}
-i && -i \\
-i && i\\
\end{pmatrix}.
\end{equation}
Note that we chose a representation for which $\det{(G)}=1$. It is then easy to see that we have 
\begin{equation}
\alpha = 2 \arccos(0) = \pi\quad\text{and}\quad \hat{n} = -\frac{1}{\sin(\pi/2)} \begin{pmatrix}
-1 /\sqrt{2}\\ 0 \\-1/\sqrt{2}
\end{pmatrix} = \begin{pmatrix}
1 /\sqrt{2}\\ 0 \\1/\sqrt{2}
\end{pmatrix}.
\end{equation}
Thus, the angular velocity simply becomes
\begin{equation}
\vec{\omega} = \frac{\pi}{\sqrt{2}\tau}\begin{pmatrix}
1\\ 0 \\1
\end{pmatrix}.
\end{equation}
From the latter we can then construct the optimal Hamiltonian
\begin{equation}
\label{hadamard-protocol}
H = \frac{\pi \hbar}{2\sqrt{2}\tau}(\sigma_x + \sigma_z)=
\frac{\pi\hbar}{2\sqrt{2}\tau} \begin{pmatrix}
0 & 1-i \\ 1+i & 0
\end{pmatrix}.
\end{equation}
It is interesting to note that this $H$ belongs to the class of Hamiltonians found in Ref.~\cite{Santos2017}, and is even identical to the ``optimal'' example studied in Ref.~\cite{deffner_energetic_2021}. However, whereas in Ref.~\cite{deffner_energetic_2021} the optimization was done somewhat heuristically, in the present analysis we have solved the optimal control problem rigorously for any single qubit quantum gates.

\section{Generalization to $N$-qubit gates}
 \label{section-4}
 
We now consider the case of a gate which operates on $N$-qubit states.  Thus, suppose that $G$ is  a $2^N \times 2^N$ unitary matrix and let $U(t)$ be the time-evolution operator, which satisfies the corresponding Schr\"odinger equation, such that $U(0) = \mathbb{I}$ and $U(\tau)=G$.  As before, we are interested in finding an optimal implementation of $G$, which minimizes the energetic cost functional \eqref{C-def}.

In complete analogy to the single qubit case, we first need to derive a version of the QSL that is governed by the eigenvalues of $U$. To this end, we will show that the eigenvalues of $U$ vary at rates which are bounded above by a function of the Hamiltonian, and therefore comparing the initial and final eigenvalues of $U$ gives a lower bound on the time-integral of that function. 

\begin{figure}
\label{arc-diagram}
\centering
\includegraphics[width=.8\textwidth]{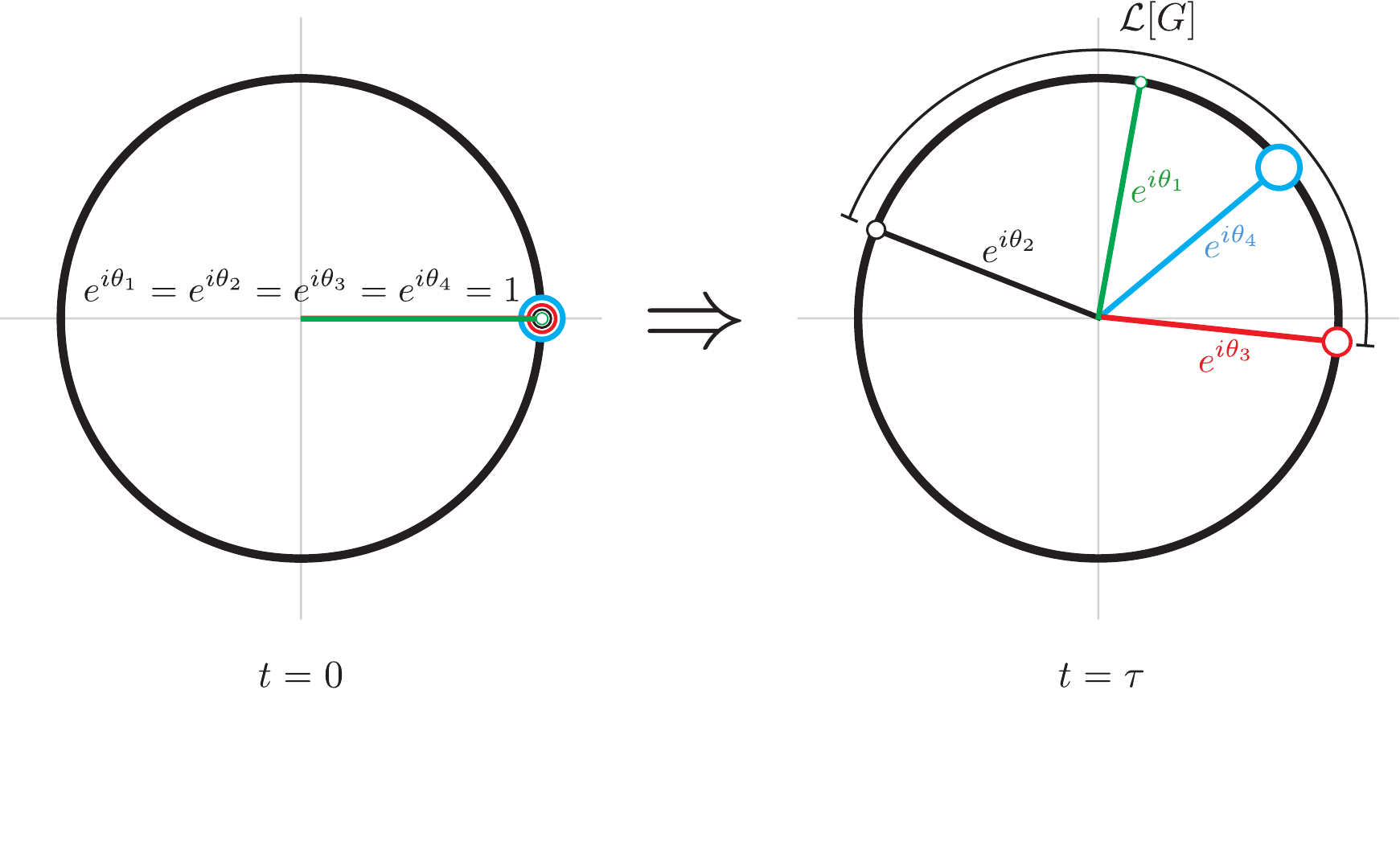}
\caption{\label{fig:superoperator}The eigenvalues of the evolution operator must spread out by an amount $L[G]$ between time $t=0$ and time $t=\tau$.}
\end{figure}

Infinitesimally,  the time evolution is given by
\begin{equation}
   \label{schrodinger-evo-2}
    U(t+dt)=U(t)-\frac{i}{\hbar}\,H(t) U(t)\,dt\,,
\end{equation}
which can be solved by means of first-order perturbation theory \cite{Messiah1962}. Thus, we write
\begin{equation}
\lambda_n(t+dt)=\lambda_n(t) -\frac{i}{\hbar}\bra{\phi_n(t)} H(t) U(t) \ket{\phi_n(t)} dt,
\end{equation}
where $\lambda_n(t)$ and $\ket{\phi_n(t)}$ are the instantaneous eigenvalues and eigenvectors of $U(t)$. Therefore, we also have
\begin{equation}
\label{lambda-ode}
\frac{\lambda_n(t+dt)-\lambda_n(t)}{dt}= -\frac{i}{\hbar}\bra{\phi_n(t)} H(t) \ket{\phi_n(t)} \lambda_n(t),
\end{equation}
Since $U(t)$ is unitary, its eigenvalues all have unit modulus,  and we can write $\lambda_n(t) = \ex{i\theta_n(t)}$. Inserting this into Eq. ~\eqref{lambda-ode} and taking the limit as $dt\to 0$ gives
\begin{equation}
\label{theta-ode}
\frac{d\theta_n}{dt} = -\frac{1}{\hbar}\bra{\phi_n(t)} H(t) \ket{\phi_n(t)}.
\end{equation}
The form of Eq. (\ref{theta-ode}) is appealing, although it suffers from the issue that the derivative of $\theta_n$ is undefined when $\theta_n$ crosses $2 \pi$ and jumps back to zero. To avoid this issue, we define a closely related\footnote{The quantity $\omega_n$ can be interpreted as the speed at which the eigenvalue $\lambda_n$ travels along the edge of the unit circle. It differs from $d\theta_n/dt$ only because the derivative of $\theta_n$ is sometimes undefined.} quantity $\omega_n$ as
\begin{equation}
\label{omega-def-2}
\omega_n = -\frac{i}{\lambda_n(t)}\frac{d \lambda_n}{dt}.
\end{equation}
It is easy to see that $\omega_n$ has the beneficial qualities that it is always defined and it is equal to the time-derivative of $\theta_n$ whenever the latter exists. From Eq. ~\eqref{lambda-ode} we then obtain
\begin{equation}
\label{omega-eq}
\omega_n = -\frac{1}{\hbar}\bra{\phi_n(t)} H(t) \ket{\phi_n(t)}.
\end{equation}
From the above it is apparent that for all $n$,
\begin{equation}
\label{omega-bound}
-E_{N-1} \leq \hbar \omega_n \leq- E_{0}.
\end{equation}
It will prove convenient to now define a superoperator $L$ called the ``shortest covering arc length''.  Given a unitary operator $X$, $L[X]$ is defined to be the length of the shortest arc on the unit circle which includes all the eigenvalues of $X$, see Fig.~\ref{fig:superoperator}.  It is clear that for a finite dimensional operator $U$, the shortest covering arc must terminate on two of its eigenvalues $\lambda_\ell$ and $\lambda_k$ for some $\ell$ and $k$. Therefore
\begin{equation}
 \label{arc-length-rate}
 \left|\frac{d}{dt}L[U(t)]\right| = |\omega_k(t) - \omega_\ell(t) |\,.
 \end{equation}

From Eq. ~\eqref{omega-bound} we then observe that
\begin{equation}
\label{arc-length-rate-bound}
 \left|\frac{d}{dt}L[U(t)]\right| \leq \frac{E_{N-1}(t)}{\hbar} - \frac{E_0(t)}{\hbar}.
\end{equation}
Recall that $U(\tau)=G$ and $U(0)=\mathbb{I}$. It is evident that $L[\mathbb{I}]=0$ since all of the eigenvalues of the identity coincide at $1$. Finally, using the fundamental theorem of calculus  we obtain
\begin{equation}
\label{arc-volume-bound}
 \int_0^\tau dt \left( E_{N-1}(t) - E_0(t)\right) \geq \hbar L[G].
\end{equation}
This takes a particularly simple expression when put in terms of the energy-time phase volume defined earlier, and we can write $ A \geq \hbar L[G]$. Note that 
\begin{equation}
\label{cont-A-H}
\frac{1}{2}(E_{N-1}-E_0) \leq \|H\| \leq E_{N-1} - E_0\,,
\end{equation}
and hence $A/2 \leq C \leq A$. In conclusion, we have
\begin{equation}
\label{energy-cost-gate}
C=\int_0^\tau dt\, \|H(t)\| \geq \hbar L[G]/2
\end{equation}
Equation~\eqref{energy-cost-gate} is the minimum energetic cost required to implement a quantum gate $G$, and thus it is a direct generalization of Eq.~\eqref{1-qubit-min-cost} to $N$-qubit gates.

\subsection{Optimal protocol}

In complete analogy to before, we now proceed to find the optimal implementation of the gate $G$, which attains the lower bound in Eq. ~\eqref{arc-volume-bound}.  We begin by assuming that for any gate $G$, there is some constant protocol which is optimal, as was the case in the single qubit problem. That is, there is a constant Hamiltonian $H$ such that $\ex{-i H \tau/\hbar} = G$, for which we have
\begin{equation}
\int_0^\tau dt\,( E_{N-1}(t) - E_0(t)) =\hbar L[G]\,.
\end{equation}
Formally, $H$ can always be found by taking the matrix logarithm of $G$, $H = i\hbar\ln(G)/\tau$.  As before, all the angles $\theta_i$ are chosen to be between $-\pi$ and $\pi$. Also note that since $G$ may be multiplied by an arbitrary phase shift, we can assume that the largest and smallest angles are chosen such that
\begin{equation}
\theta_{N-1} - \theta_0 = L[G],
\end{equation}
and such that $\theta_{0}=-\theta_{N-1}$. The eigenvalues of $H$ are then simply $E_n = \hbar\theta_n/\tau$. It is easy to see that  the energy-time phase volume reads $A = \hbar (\theta_{N-1}-\theta_0)$, and that the corresponding energetic cost becomes $C = \hbar L[G]/2$.  Thus, as was the case for single qubit gates, optimal implementations of arbitrary $N$-qubit gates are given by constant Hamiltonians.

\subsection{Nonuniqueness of the optimal protocol}

The immediate question arises, whether the constant protocol is the unique optimal implementation, or whether there exist other protocols with the same, minimal energetic cost. We will show now that there is is actually  an infinite family of optimal protocols, see also Refs.~\cite{Acconcia2015PRE,Acconcia2015PREq,deffner_energetic_2021} for similar conclusions.

Consider a function $f(t)$, which is everywhere nonnegative and satisfies $\int_0^\tau  dt f(t) = \tau$. Then let the Hamiltonian be given as a function of time by $H'(t) = f(t) i\hbar \ln (G)/\tau$ where the logarithm is chosen as above.  Then the evolution operator reads
\begin{equation}
U(\tau) = \exp \left( \ln(G)\, \frac{1}{\tau} \int_0^\tau dt\, f(t) \right) = G,
\end{equation}
One easily convinces oneself that $H'(t)$ is an allowed implementation.

To see that this protocol is also optimal, note that since $H'(t) = f(t) H$ and $f(t)$ is nonnegative, we have $\|H'(t)\| = f(t) \|H\|$
and, hence,  $\int_0^\tau dt \|H'(t)\| = \|H\| \int_0^\tau dt f(t)$. Thus, we immediately have
\begin{equation}
\int_0^\tau dt\, \|H'(t)\| = \|H\| \int_0^\tau dt\, f(t) = \int_0^\tau dt\, \|H\|,
\end{equation}
which shows that $H'(t)$ is optimal.

\section{Higher order cost functionals}
\label{section-5}

Having found an infinite family of implementations that are optimal with respect to the energetic cost based on the operator norm \eqref{C-def} raises the question whether working with other norms would given more insight or restrictions. Thus, we conclude the analysis with generalizations of our results to the Lebesgue $p$-norm and the Schatten $p$-norm for arbitrary $p$.

\subsection{Lebesgue norm cost functional}

We start with the Lebesgue $p$-norm, and consider the modified cost functional
\begin{equation}
\label{lebesgue-cost}
C_{p}[H] = \left(\int_0^\tau dt\,\|H(t)\|^p\right)^{1/p} .
\end{equation}
Given $p\geq 1$, H\"older's inequality states that
\begin{equation}
\label{Holder}
\int_0^\tau dt\,|f(t)g(t)| \leq \left( \int_0^\tau dt\, |f(t)|^p\right)^{1/p}\left( \int_0^\tau dt |g(t)|^q\right)^{1/q},
\end{equation}
where $q=1/(1-p^{-1})$.  Setting $g(t)=1$, $f(t)=\|H(t)\|$ and $p=1$ we obtain
\begin{equation}
\int_0^\tau dt\,\|H(t)\| \leq \left( \int_0^\tau dt \|H(t)\|^p\right)^{1/p}\tau^{1/q},
\end{equation}
which leads to a lower bound on $C_p$,
\begin{equation}
C_p[H]\geq \tau^{-1/q}\, C_1 [H].
\end{equation}
Then using the definition of $q$ and the lower bound on $C_1\geq \hbar L[G]/2$ obtained above in Eq.~\eqref{energy-cost-gate}, we can write
\begin{equation}
\label{lebesgue-bound}
C_p[H] \geq  \frac{\hbar L[G]}{2}\,\tau^{(1-p)/p}.
\end{equation}
H\"older's inequality becomes an equality iff the functions $f$ and $g$ are linearly dependent. This means that the minimal cost is attained only if $\|H(t)\|$ is constant. Thus,  working with the Lebesgue $p$-norm identifies the constant protocol as the \emph{unique} optimal implementation. It is interesting to note that Eq. ~\eqref{lebesgue-bound} gives a lower bound on the cost which depends on protocol duration, unlike Eq. (\ref{energy-cost-gate}) which did not depend on the protocol duration.

An interesting case is $p=2$ for a single qubit, for which we can write the cost functional as
\begin{equation}
\label{lebesgue-cost-qubit}
C_2[H]=\sqrt{\int_0^\tau dt\,|\vec{\omega}(t)|^2}\,.
\end{equation}
Equation~\eqref{lebesgue-cost-qubit} is proportional to the square root of the dynamical action for a rigid rotating sphere with no external potential. This implies that minimizing the cost $C_2$ is equivalent to solving for the evolution of a rigid rotating sphere between time $t=0$ and time $t=\tau$, subject to the constraint that the initial and final orientations are specified.  Thus, it is no surprise that a constant-axis rotation with constant angular velocity gives an optimal solution.

\subsection{Schatten norm cost functional}

Finally, we consider a generalized cost which is given by
\begin{equation}
\label{schatten-cost}
C_p[H]=\int_0^\tau dt\,\|H\|_p,
\end{equation}
where $\|H\|_p$ is the Schatten $p$-norm \eqref{schatten-p}.  In this case, it will prove instructive to demand that the gate $G$ is implemented exactly, without allowing an arbitrary phase factor. That is, we require that $U(\tau) = G$. As before, we parameterize the eigenvalues of $G$ as $\lambda_j=\ex{i\theta_j}$, with $\theta_j$ between $-\pi$ and $\pi$. Now using the triangle inequality for the vector $p$-norm gives
\begin{equation}
\label{triangle-omega}
\int_0^\tau dt\, |\vec{\omega}(t)|_p\geq |\vec{\theta}(\tau)|_p,
\end{equation}
where $|\vec{\theta}|_p = (\sum_j |\theta_j|^p)^{1/p}$ and $|\vec{\omega}|_p = (\sum_j |\omega_j|^p)^{1/p}$. Using Eq. (\ref{omega-eq})  we then obtain
\begin{equation}
|\vec{\omega}|_p =\frac{1}{\hbar}\left(\sum_j |\braket{\phi_j}{H|\phi_j}|^p \right)^{1/p}\,,
\end{equation}
and since $|\braket{\phi_j}{H|\phi_j}|^p \leq \braket{\phi_j}{|H|^p|\phi_j}$, we have
\begin{equation}
\label{eq:p-norms}
|\vec{\omega}|_p \leq\frac{1}{\hbar}\left(\sum_j \braket{\phi_j}{|H|^p|\phi_j} \right)^{1/p} = \frac{1}{\hbar}\,\tr{(|H|^p)^{1/p}}.
\end{equation}
The right hand side of Eq.~\eqref{eq:p-norms} is simply an equivalent expression for the Schatten $p$-norm. Hence, $ |\vec{\omega}|_p \leq \|H\|_p/\hbar$, and we obtain  a lower bound on the cost
\begin{equation}
\label{schatten-min-1}
\int_0^\tau dt\,\|H(t)\|_p \geq \hbar\, | \vec{\theta}|_p.
\end{equation}
Solving the optimal control problem for Eq.~\eqref{schatten-min-1} is significantly more involved, which is why we leave this issue for future work.

\section{Concluding Remarks}
\label{section-6}
The results in this paper build on recent progress in the field of QSL to give physical constraints on the best-case performa{}nce of a quantum computer. Specifically we have found the maximal energy efficiency of quantum gates, and the protocol which attains that maximum. While energy-efficiency is not as commonly discussed as certain other notions of performance (e.g. time-complexity of algorithms), we believe that the energy efficiency is an important figure of merit which deserves further study.

Historically, research on maximal energy efficiency of physical systems has a rich tradition, perhaps the most celebrated example being Carnot's foundational contribution to thermodynamics.  Since Carnot not only found the maximal efficiency of a heat engine, but also provided a protocol which attains this efficiency, his work served as a template for both physicists and engineers. Ultimately this work made it possible to refine the creations of the early industrial era into the sophisticated machines which are available today.

A full thermodynamic understanding of quantum computing may prove similarly revolutionary. In order to achieve such an understanding, it will be necessary to combine bounds on best-case performance (such as QSLs) with a quantum treatment of work and heat. For this reason, one of the more pressing questions related to this work is: ``How exactly is the energetic cost functional related to the heat and work inputs and outputs of a quantum system?" In addition to this question, a number of other questions remain to be answered. An important consideration which we have neglected in this paper is the cost of separating the system from the environment. Similarly, the use of error-correcting protocols likely increases the minimal cost needed to implement a quantum gate. These additional costs may represent a significant portion of the actual energy needed to operate a quantum computer in practice, which we leave to be investigated in future work. 

\invisiblesection{Acknowledgments}
\section*{Acknowledgments}

M.A. gratefully acknowledges support from Harry Shaw of NASA Goddard Space Flight Center and Kenneth Cohen of Peraton. 

\section*{Appendix}

\renewcommand{\theequation}{A.\arabic{equation}}

\setcounter{equation}{0}

\appendix
\subsection{Proof that identity term only contributes an overall phase}
\label{appendix-1}

Recall that the differential equation for the evolution operator is
\begin{equation}
U(t+dt) = \left(\mathbb{I} - \frac{i}{\hbar}H(t)dt\right)U(t)
\end{equation}
Putting in the Hamiltonian $H = u_0(t) \mathbb{I} + \vec{u}(t) \cdot \vec{\sigma}$ gives
\begin{equation}
U(t+dt)=\left[\mathbb{I} -  \frac{i}{\hbar}(u_0(t) \mathbb{I} +\vec{u}(t) \cdot \vec{\sigma})dt\right]U(t) dt.
\end{equation}
Separating the time interval into $M$ discrete steps, we then write the final evolution operator $U(\tau)$ as a product
\begin{equation}
U(\tau)= \prod_{j=1}^M \left[\mathbb{I} -  \frac{i}{\hbar}(u_0(t_j) \mathbb{I} +\vec{u}(t_j) \cdot \vec{\sigma})dt\right].
\end{equation}
Note that to first order in $dt$,
\begin{equation}
\left(\mathbb{I} - \frac{i}{\hbar}u_0 \mathbb{I}dt\right)\left(\mathbb{I}-\frac{i}{\hbar} \vec{u} \cdot \vec{\sigma}dt\right) = \mathbb{I} - \frac{i}{\hbar}(u_0 \mathbb{I} +\vec{u}\cdot \vec{\sigma})dt.
\end{equation}
Therefore we can factor each term in the product in this way, to obtain
\begin{equation}
U(\tau)= \prod_{j=1}^M \left(\mathbb{I} - \frac{i}{\hbar}u_0(t_j) \mathbb{I}dt\right)\left(\mathbb{I}-\frac{i}{\hbar} \vec{u}(t_j) \cdot \vec{\sigma}dt\right).
\end{equation}
Since the first factor is a constant multiplied by the identity, it commutes with any operator. Therefore
\begin{equation}
U(\tau)= \prod_{j=1}^M \left(\mathbb{I}-\frac{i}{\hbar} \vec{u}(t_j) \cdot \vec{\sigma}dt\right)\prod_{j=1}^M \left(\mathbb{I} - \frac{i}{\hbar}u_0(t_j) \mathbb{I}dt\right).
\end{equation}
We then recognize the limit of the second product as a scalar exponential multiplied by the identity operator
\begin{equation}
\prod_{j=1}^M \left(\mathbb{I} - \frac{i}{\hbar}u_0(t_j) \mathbb{I}dt\right) = \mathbb{I} \exp \left(-\frac{i}{\hbar} \int_0^\tau dt u_0(t)\right).
\end{equation}
Since the other factors in the evolution operator do not depend on $u_0$, we have shown that $u_0$ only contributes an overall phase shift.

\subsection{Proof that variable-axis protocols are non-optimal}
\label{appendix-2}

It was shown in Sec.~\ref{section-3} that there is an optimal protocol for a single qubit gate which rotates the Bloch sphere about a fixed axis. We now show that protocols which rotate the Bloch sphere about an axis which is not fixed are not optimal. Suppose that $\vec{v}_0$ is a vector perpendicular to $\hat{n}$, and that $\vec{v}(t)$ is the trajectory of the point which is initially at $\vec{v}_0$. Let $\vec{\omega}_\parallel(t)$ be the component of $\vec{\omega}$ which is parallel $\vec{v}(t)$ and $\vec{\omega}_\perp(t)$ be the component of $\vec{\omega}$ which is perpendicular to $\vec{v}(t)$. Note that 
\begin{equation}
\left| \frac{d\vec{v}}{dt} \right| = |\vec{\omega}_\perp|,
\end{equation}
so if $\omega_\parallel$ is not zero almost everywhere then
\begin{equation}
\int_0^\tau dt|\vec{\omega}|=\int_0^\tau dt \sqrt{|\vec\omega_\parallel|^2+|\vec\omega_\perp|^2}>\int_0^\tau dt\sqrt{|\vec\omega_\perp|^2} = \int_0^\tau dt \left| \frac{d\vec{v}}{dt} \right|
\end{equation}
However it was already shown in Sec.~\ref{section-3} that 
\begin{equation}
\int_0^\tau dt \left| \frac{d\vec{v}}{dt} \right| \geq \alpha,
\end{equation}
which implies
\begin{equation}
\int_0^\tau dt\,|\vec{\omega}|dt>\int_0^\tau dt \left| \frac{d\vec{v}}{dt} \right|\geq \alpha \,.
\end{equation}
 Therefore this protocol cannot be optimal. Because the above argument applies for any choice of $\vec{v}_0$ which is perpendicular to $\hat{n}$, the protocol cannot be optimal unless $\vec{\omega}$ is always parallel to $\hat{n}$, so the axis of rotation must be fixed.

\section*{References}

\bibliographystyle{iopart-num}
\bibliography{ec.bib} 

\end{document}